\documentclass[amsfonts,amssymb]{revtex4}

\usepackage{indentfirst}
\usepackage[reqno,intlimits]{amsmath}
\usepackage[polish,english]{babel}

\newcommand{\ud}{\mathrm{d}}

\renewcommand{\L}{\Omega_{\Lambda 0}}
\renewcommand{\k}{\Omega_{K0}}
\newcommand{\m}{\Omega_{m0}}
\renewcommand{\r}{\Omega_{r0}}

\newtheorem{theorem}{Theorem}
\newtheorem{lemma}{Lemma}
\newtheorem{definition}{Definition}

\begin{document}

\title{Non-integrability of density perturbations in the FRW universe}

\author{Tomasz Stachowiak}
\email{toms@oa.uj.edu.pl}
\affiliation{\it Astronomical Observatory, Jagiellonian University,
ul. Orla 171, 30-244 Krak\'ow, Poland}

\author{Marek Szyd\lpb owski}
\email{uoszydlo@cyf-kr.edu.pl}
\affiliation{\it Astronomical Observatory, Jagiellonian University,
ul. Orla 171, 30-244 Krak\'ow, Poland}

\author{Andrzej J. Maciejewski}
\email{maciejka@astro.ia.uz.zgora.pl}
\affiliation{\it Institute of Astronomy, University of Zielona G\'ora, ul.
Podg\'orna 50, 65-246 Zielona G\'ora, Poland}

\begin{abstract}
We investigate the evolution equation of linear density perturbations in
the Friedmann-Robertson-Walker universe with matter, radiation and the
cosmological constant. The concept of solvability by quadratures is defined
and used to prove that there are no ``closed form'' solutions except for the
known Chernin, Heath, Meszaros and simple degenerate ones. The analysis is
performed applying Kovacic's algorithm. The possibility of the existence of
other, more general solutions involving special functions is also investigated.
\end{abstract}

\maketitle

\section{Introduction}

This paper is an attempt to apply the methods of differential algebra, and the
differential Galois theory in particular, to a problem of cosmology. Although
the considerations are mostly mathematical, the problem itself, and its
solutions, are of rather practical interest. Linear perturbations of the
Einstein equations in many forms are investigated due to their direct relation
to such practical questions as the formation of galaxies or the CMB
inhomogeneities. We, however, enter the physical domain only as the source of a
theoretical problem, on which we concentrate.

The result itself is of negative nature, or, in other words, it makes any
further investigation of this kind unnecessary. All the known solutions are
given, together with the conditions for their validity. As no new ones can
exist, this closes and completes the analysis of the given equation.

Of course, that is not to say that apart from those special cases nothing can
be said about the behaviour of the solution. It is only to some extent that
differential algebra can make exact the intuitive concepts of ``being possible
to solve'' or ``expressible in a closed form''. The definition of
non-integrability employed here is but one of many which were born as
classical mechanics evolved. The Liouville's theorem implies that enough first
integrals might yield a complete solution of a dynamical system, and
accordingly many criteria regarding the existence of certain classes of first
integrals were developed. The first achievements were those of Kovalevskaya
and Lyapunov, greatly improved only recently by Ziglin, Morales and Ramis.
The Galois theory used here, can also be applied to prove non-existence of
meromorphic first integrals in more complex systems. Paradoxically here, we are
presented with an equation simple enough not to allow for the application of
those advanced methods. Thus, only a small part of the theory is put into
practise and explained here. For a complete bibliography and exposition see for
example \cite{Morales}.

The paper is organised as follows. In section 2 we derive the equation in
question in a non-standard but intuitively clear way. The next two
sections describe the concept of Liouvillian solutions, integrability and give
the basic criteria, which we proceed to use in section 5. We also
investigate the possibility of solving the problem by a combination of some
special and Liouvillian functions, and give the theorem which is the main
result in section 6. Finally, conclusions and finishing remarks are
given in section 7.

\section{Density Perturbation Equation}

We will be considering the Friedmann-Robertson-Walker universe given by the
metric
\begin{equation}
    \ud s^2 = c^2 \ud t^2 - a(t)^2 \left[\frac{\ud r^2}{1-Kr^2}+
        r^2\ud\Omega^2\right],
\end{equation}
where $K$ is the curvature index, $\ud\Omega^2$ the distance element on a
two-sphere. The universe will be filled with radiation and baryonic matter
characterised by their pressures and densities $p$ and $\rho$.
A non-zero cosmological constant's effect will also be considered.

The Einstein equations for this model give
\begin{equation}
\left\{
\begin{aligned}
    \left(\frac{\dot{a}}{a}\right)^2 &= \frac{8\pi\boldsymbol{G}}{3}\rho +
    \frac{\Lambda}{3} - \frac{K}{a^2} \\
    \frac{\ddot{a}}{a} &= -\frac{4\pi\boldsymbol{G}}{3}(3p+\rho) +
    \frac{\Lambda}{3},
\end{aligned}
\right. \label{fried}
\end{equation}
where $\boldsymbol{G}$, and $\Lambda$ are the gravitational and cosmological
constants respectively. The conservation equation reduces to
\begin{equation}
    c^2\dot\rho = -3 H (p+\rho c^2),
\end{equation}
with the dot representing the time derivative. This can be expressed as the
following transformation laws for matter and radiation respectively:
\begin{equation}
\begin{aligned}
    p_m &= 0 &\Rightarrow\quad
    \frac{\rho_m}{\rho_{m0}} = \left(\frac{a_0}{a}\right)^{3},\\
    p_r &= \tfrac13\rho_r &\Rightarrow\quad
    \frac{\rho_r}{\rho_{r0}} = \left(\frac{a_0}{a}\right)^{4}.\\
\end{aligned} \label{scale}
\end{equation}
$\rho_{m0}$ and $\rho_{r0}$ are the values of the densities for the moment
when $a=a_0$, which can be chosen as the present day.

The fluctuation is introduced by the means of the scale factor
\begin{equation}
    a = \tilde{a}(1+y),
\end{equation}
and $\tilde{a}$ is the solution of the original equations (\ref{fried}).
As it is a matter perturbation only, we have
\begin{align*}
    \rho_r &= \tilde\rho_r,\\
    \rho_m &= \tilde\rho_m \left(\frac{a}{\tilde{a}}\right)^{-3}\\
        &= \tilde\rho_m(1+y)^{-3}\\
        &= \tilde\rho_m(1-3y).
\end{align*}
Where we use the scaling law (\ref{scale}) and linearise the problem.
Substituting this into the second of the equations (\ref{fried}), we obtain
\begin{equation}
    \ddot{\tilde{a}}(1+y)+2\dot{\tilde{a}}\dot{y} = 
    -\tfrac{8\pi\boldsymbol{G}}{3}\tilde{a}(1+y)\rho_r
    -\tfrac{4\pi\boldsymbol{G}}{3}\tilde{a}(1+y)\tilde\rho_m(1-3y)
    +\tfrac13\Lambda\tilde{a}(1+y),
\end{equation}
which, after substituting the equation satisfied by the unperturbed $\tilde{a}$,
simplifies to
\begin{equation}
    \ddot{y} + 2 H \dot{y} - 4\pi\boldsymbol{G}\tilde{\rho}_m y = 0,
        \label{main_eq1}
\end{equation}
with the Hubble ``constant''
\begin{equation}
    H := \frac{\dot{\tilde{a}}}{\tilde{a}}.
\end{equation}

This kind of perturbation is of the scalar type, i.e. constructed from a single
function $y$. Here, we take it to depend on time only, although in general it
could also involve spatial variables. This case could then be thought of as the
zeroth order coefficient in the expansion of $y(t,r,\theta,\varphi)$ in terms
of eigenfunctions of the spatial Laplace operator. Another generalisation would
be to consider the vector and tensor type perturbations (see \cite{Stewart} for
details of the decomposition).
As it turns out, vector perturbations also admit quite general exact solutions
\cite{LKB}.

In order to simplify the equation (\ref{main_eq1}) fully, that is, bring it to
the linear form with rational coefficients, we choose new variables
\begin{equation}
    x := \frac{\tilde{a}}{\tilde{a}_0},\; u := H_0 t,
\end{equation}
and constant density parameters:
\begin{align}
    \r &:= \frac{8\pi \boldsymbol{G} \rho_{r0}}{3H_0^2}, & 
    \m &:= \frac{8\pi \boldsymbol{G} \rho_{m0}}{3H_0^2}\\
    \k &:= -\frac{c^2 K}{\tilde{a}_0^2 H_0^2}, &
    \L &:= \frac{c^2 \Lambda}{3 H_0^2}
\end{align}
which allow us to rewrite the main equation as
\begin{equation}
    \left(\frac{\ud x}{\ud u}\right)^2 \frac{\ud^2y}{\ud x^2} +
    \left[\frac{\ud^2 x}{\ud u^2} + \frac{2}{x}\left(\frac{\ud x}{\ud u}\right)^2\right] \frac{\ud y}{\ud x} -
    \tfrac32\m \frac{1}{x^3}y = 0,
\end{equation}
where the perturbation $y$ is now considered as a function of $x$. Finally,
using the first of the equations (\ref{fried}), which in the new variables
reads
\begin{equation}
    x^2\left(\frac{\ud x}{\ud u}\right)^2 =
        \L x^4 + \k x^2 + \m x + \r =: W(x), \label{elliptic}
\end{equation}
we can eliminate the
derivatives with respect to $u$, and denoting the differentiation with respect
to $x$ with a prime, we get
\begin{equation}
    x(\L x^4 + \k x^2 + \m x + \r)y'' 
    + (3\L x^4 + 2\k x^2 + \tfrac32\m x + \r)y' - \tfrac32\m y = 0. \label{main_eq2}
\end{equation}

As follows from the definitions, $\k$ and $\L$ are of arbitrary signs, while
$\r$ and $\m$ are non-negative. We take $\m$ to be strictly positive, though,
because of the nature of the examined perturbations.

We also introduce the functions $p(x)$ and $q(x)$ related to the above equation
in the following form:
\begin{equation}
    y''(x) + p(x) y'(x) + q(x) y(x) = 0. \label{rational_eq}
\end{equation}
They are
\begin{equation}
    p(x) = \frac{6\L x^4 + 4\k x^2 + 3\m x + 2\r}
    {2x(\L x^4 + \k x^2 + \m x + \r)},\mathrm{and}\;\;\;
    q(x) = \frac{-3\m}{2x(\L x^4 + \k x^2 + \m x + \r)}.
\end{equation}

Having obtained the solutions as functions of $x$ it is straightforward to
express them as functions of the cosmological or conformal time, since $x(u)$
satisfies equation (\ref{elliptic}), and is therefore expressible in terms of
the elliptic functions. The exact formulae can be found for example in
\cite{phantom}.

\section{Liouvillian Solutions}

Equation (\ref{main_eq2}) is, in general, a
Fuchsian one, and the solutions can be found by means of series. In particular,
in vicinities of the singular points
\begin{equation}
    y_{\pm}(x) = (x-x_0)^{\alpha_{\pm}}
    \left(1+\sum_{n=1}^{\infty}c_n (x-x_0)^n\right),
\end{equation}
where $\alpha_{\pm}$ are the characteristic exponents at the point $x_0$. Such
solutions are local, and the area of convergence around a given point is
restricted by the remaining singular points.
However, there are some special cases in which the solutions can be
expressed by means of known special functions, and become global.
It is then much easier to
investigate and understand their behaviour. Thus, we are lead to the natural
question of existence of such ``simple'' solutions. In this section we give a
short description of a class of functions, which could here be called
closed-form solutions.

It is natural to start seeking for the solutions in a class of functions to
which the given equation's coefficients belong, but such a set proves
to be insufficient in most cases. As we are dealing with an equation whose
coefficients are rational functions, the first choice is the field
$\mathbb{C}(x)$ -- rational functions with complex numbers as the field of
constants. Or, in the language of differential algebra,
$(\mathbb{C}(x),\partial)$ - the above field equipped with a suitable
derivation operation, which, in this case, coincides with the usual, complex
one, and will be denoted by a prime throughout this section.

This field is much too small, of course, and we are soon forced to
extend it, introducing new elements. Just, as $\mathbb{C}(x)$ is an
extension of $\mathbb{C}$ obtained by adjoining an indeterminate variable $x$,
our new fields will be $\mathbb{C}(x)$ to which new functions are added.
Naturally, the derivation on the extended field must coincide with the
subfield's derivation when restricted to it (that is essentially what a
differential extension is).

To keep the new functions relatively simple, so that the usual notion of
``solvability by quadratures'' could still be applied, the new elements are
restricted to three classes.
\begin{definition}
    For a differential field extension $F\subset G$, an element $a\in F$ is:
\begin{description}
\item[primitive] over $G$
    if $a'\in G$,
\item[exponential] over $G$
    if $a'/a\in G$, or
\item[algebraic] over $G$
    if $P(a)=0$ for some $P(x)\in G[x]$ -- the ring of polynomials with
    coefficients in $G$.
\end{description}
\end{definition}

\begin{definition}
    A field extension is called Liouvillian if it is a result of a finite
    number of extensions, each the adjunction of an algebraic, exponential,
    or primitive element.
\end{definition}
Some examples of ``new'' functions appearing in this process are radicals (for
algebraic elements), logarithms and inverse circular function (for primitives),
and trigonometric functions (for exponentials). In short, the new elements
are expressible as combinations of exponentials, integrals, and algebraic and
elementary functions.

Finally, we are able to formulate what we mean by integrability:
\begin{definition}
    A linear differential equation with coefficients in $\mathbb{C}(x)$ is
    said to posses Liouvillian solutions, or be solvable by quadratures,
    if its solutions belong to a Liouvillian extension of $\mathbb{C}(x)$.
\end{definition}

\section{Monodromy and Galois Groups}

In order to characterise a differential equation we can introduce two groups
whose invariant properties are closely connected with those of the first
integrals of the given equation.

The first, called the monodromy group $\mathcal{M}$ is associated with the
continuation of the local solutions of a linear differential equation, along
closed paths in the domain where the equation is defined. The group itself is
an image the fundamental group of that domain, and a subgroup of GL($n,\mathbb C$)
for $n$ the order of the equation. The only other fact, concerning
$\mathcal M$, that we will need here is that $\mathcal M\subset \mathcal G$ --
the differential Galois group, which we proceed to describe.

$\mathcal G$ is directly connected with the extension of the base field
$\mathbb C(x)$ to a bigger field $F$, which contains the solutions of the
considered equation. $\mathcal G$ is defined as the group of automorphisms of
$F$ that leave $\mathbb C(x)$ element-wise fixed. Carrying this information
regarding the equation, the Galois group enables to test for particular forms
of solutions. One of the fundamental properties is as follows:

\begin{lemma}
    Let $\mathcal G$ be the differential Galois group of a linear differential
    equation $w''(x)=r(x)w(x)$, $r(x)\in\mathbb C(x)$. Then it is an
    algebraic subgroup of SL$(2,\mathbb C)$, and one of the following cases
    can occur.
\begin{enumerate}
\item $\mathcal G$ is triangulisable. There is a solution of the form
$\exp\int\omega$, where $\omega\in\mathbb C(x)$.
\item $\mathcal G$ is conjugate to a subgroup of
\begin{displaymath}
D^{\dag} = \left\{
\left( \begin{array}{cc}
c_1 & 0 \\
0 & c_1^{-1}
\end{array} \right) \bigg| c_1\in \mathbb C^* \right\}
\cup \left\{
\left( \begin{array}{cc}
0 & c_2 \\
-c_2^{-1} & 0 
\end{array} \right) \bigg| c_2\in \mathbb C^* \right\},
\end{displaymath}
where $\mathbb C^*=\mathbb C \setminus\{0\}$. There is a solution of the form
$\exp\int\omega$, with $\omega$ algebraic over $\mathbb C(x)$ of degree 2.
\item $\mathcal G$ is finite. All the solutions are algebraic.
\item $\mathcal G =$ SL$(2,\mathbb C)$. The equation has no Liouvillian
solutions.
\end{enumerate} \label{3cases}
\end{lemma}

\section{Non-integrability of the Equations}

We now proceed to the direct analysis of the integrability of equation
(\ref{main_eq2}), based mainly on the Kovacic's algorithm \cite{Kovacic}.
First we consider the equation with $\L=0$, for which the calculation is not
too cumbersome, and next, we outline the reasoning for the general case.

\subsection{Models without the cosmological constant}

In order to apply the algorithm, equation (\ref{rational_eq}) is cast into
a reduced form
\begin{equation}
    v''(x)=r(x)v(x), \label{reduced}
\end{equation}
with
\begin{equation}
    r(x) = -\frac{1}{4x^2}
        -\frac{3(2\kappa x + 1)^2}{(\kappa x^2 + x + \varrho)^2}
        +\frac{4\kappa x+7}{x(\kappa x^2 + x + \varrho)}.
\end{equation}
Where we have introduced new parameters
\begin{equation*}
    \kappa := \frac{\k}{\m}, \qquad \varrho := \frac{\r}{\m}.
\end{equation*}
The reduction itself is performed by means of the following change in the
dependent variable
\begin{equation}
    y(x) = v(x) \exp\left[-\tfrac12\int_{x_0}^x p(s)\ud s\right],
\end{equation}
where the constant $x_0$ is arbitrary. This transformation does not change the
class of solutions we are interested in, as it uses the ``admissible''
operations only.

In the general case, there are two distinct roots of the polynomial $W(x)$
in equation (\ref{main_eq2}) and they are both non-zero.
The degenerate cases will be treated separately.

Looking for the local solutions around $x=0$, which is a regular singular
point of the equation with both characteristic exponents equal to $\frac12$,
we find
\begin{equation}
\begin{aligned}
    v_1(x) &= x^{1/2} w_1(x), \\
    v_2(x) &= \ln(x) v_1(x) + x^{1/2} w_2(x),
\end{aligned}
\end{equation}
and $w_1$ and $w_2$ are holomorphic at $x=0$. Using these,
it is now possible to obtain an element of the monodromy group, by continuation
of the fundamental solution matrix along a small closed path around $x=0$.
\begin{equation}
    \left(
    \begin{array}{cc}
    v_1 & v_2 \\
    v_1' & v_2'
    \end{array}
    \right) \longrightarrow
    \left(
    \begin{array}{cc}
    -1 & -2\pi i \\
    0 & -1
    \end{array}
    \right) \left(
    \begin{array}{cc}
    v_1 & v_2 \\
    v_1' & v_2'
    \end{array}
    \right)
\end{equation}

Since $\mathcal M \subset\mathcal G$, the Galois Group cannot be $D^{\dag}$, as
that group does not contain non-diagonalisable matrices. Also, a triangular,
non-diagonal matrix cannot generate a finite group, so that $\mathcal G$ itself
cannot be finite. We have thus excluded cases 2 and 3 of lemma~\ref{3cases}.

Case 1 can also be easily excluded applying the aforementioned algorithm.
In order for the solution to be of the form
\begin{equation}
    P(x)\exp\left[\int\omega(x)\ud x\right],
\end{equation}
with a monic $P(x)\in\mathbb C[x]$, and $\omega(x)\in\mathbb C(x)$, the
following equation must be satisfied:
\begin{equation}
    P''(x)+2\omega(x)P'(x)+[\omega'(x)+\omega(x)^2-r(x)]P(x)\equiv0,
    \label{p_om_eq}
\end{equation}
for $P(x)$ and $\omega(x)$ found according to the algorithm, as follows.
First, we define a set of auxiliary quantities:
\begin{equation}
    \alpha_c^{\pm} = \tfrac12 \pm \tfrac12\sqrt{1+4b},
\end{equation}
where $c\in\Gamma\cup\{\infty\}$ -- the set of all the finite poles of $r(x)$
including infinity, and $b$ is the coefficient of $(x-c)^{-2}$ in the Laurent
expansion of $r(x)$ around a given point $x=c$. Next, for all possible
combinations of signs $s(\infty), s(c)$ we compute the possible degrees of
$P(x)$
as
\begin{equation}
    d = \alpha_{\infty}^{s(\infty)} - \sum_{c\in\Gamma} \alpha_c^{s(c)},
\end{equation}
and, for the same combinations of signs $s(c)$, the respective possible
$\omega$
\begin{equation}
    \omega(x) = \sum_{c\in\Gamma}\frac{\alpha_c^{s(c)}}{x-c}.
\end{equation}
Inserting the functions just obtained into equation
(\ref{p_om_eq}) we are left with a system of linear equations determining the
unknown coefficients of $P(x)$. In this particular case, the only non-negative
value of $d$ is 0, so that $P(x)=1$, and the system is reduced to
\begin{equation}
    \frac{3}{2x(\kappa x^2 + x +\varrho)}\equiv 0,    
\end{equation}
which cannot hold. The first case of
lemma~\ref{3cases} is also excluded, which means that $\mathcal G$ is
SL$(2,\mathbb C)$, and the equation has no Liouvillian solutions in the
general case.

When we admit a double non-zero root of $W(x)$, or in other words, when
\begin{equation}
    \m=2\sqrt{\r}\left(1-\sqrt{\r}\right), \label{GZ_cond}
\end{equation}
we have the well known Chernin solution \cite{Chernin}
\begin{equation}
    y(x) = {}_{2}F_{1}\left(-i\sqrt3,i\sqrt3,1;\frac{x}{x+2\r}\right).
    \label{Chernin_sol}
\end{equation}

In the current notation it means that $4\varrho\kappa=1$, and that the root
itself is $x_0=-2\varrho$. As equation (\ref{rational_eq}) now has three
regular singular points: 0, $x_0$ and $\infty$, it becomes a Riemann
P-equation. The complete set of solutions is denoted by
\begin{equation}
    y(x) = P\left\{
        \begin{array}{cccc}
            \; 0 \;&\; -2\varrho \;&\; \infty \; &\\
            \; 0 \;&\; -i\sqrt3  \;&\; 0      \; &\; x \;\\
            \; 0 \;&\; i\sqrt3   \;&\; 1      \; &
        \end{array} \right\}.
\end{equation}
Such a P-function is not Liouvillian either, as can be immediately checked
using Kimura's Theorem \cite{Kimura}.

Another degenerate subcase occurs when $\varrho=0$, and $\kappa\ne 0$, so
that 0 becomes a root of $W(x)$ and there is another non-zero root
$x=-\frac{1}{\kappa}$. Like before the solution is a P-function
\begin{equation}
    y(x) = P\left\{
        \begin{array}{cccc}
            \; 0         \;&\; -\tfrac{1}{\kappa} \;&\; \infty \; &\\
            \; -\tfrac32 \;&\; 0                  \;&\; 0      \; &\; x \;\\
            \; 1         \;&\; \tfrac12           \;&\; 1      \; &
        \end{array} \right\}.
\end{equation}
This time however, it is a Liouvillian solution because we can express the
above symbol as two base solutions:
\begin{equation}
\begin{aligned}
    y_1(x) &= x^{-3/2}\sqrt{\m+\k x},\\
    y_2(x) &= x\,{}_2F_1\left(1,2,\tfrac72;-\frac{\k}{\m}x\right),
\end{aligned} \label{Heath_0}
\end{equation}
where ${}_2F_1$ is the Gauss hypergeometric function.

The next subcase is $\kappa = 0$ and only two simple roots remain: 0 and
$-\varrho$. We accordingly get
\begin{equation}
    y(x) = P\left\{
        \begin{array}{cccc}
            \; 0 \;&\; -\varrho \;&\; \infty   \; &\\
            \; 0 \;&\; 0        \;&\; -1       \; &\; x \;\\
            \; 0 \;&\; \tfrac12 \;&\; \tfrac32 \; &
        \end{array} \right\}.
\end{equation}
As before, this is a Liouvillian function, according to Kimura's theorem, and
it can be rewritten as the following two independent solutions:
\begin{equation}
\begin{aligned}
    y_1(x) &= 1 + \frac{3\m}{2\r}x,\\ 
    y_2(x) &= \sqrt{\r+\m x}\,
        {}_2F_1\left(2,-\tfrac12,\tfrac32;1+\frac{\m}{\r} x\right),
\end{aligned} \label{flat_sol}
\end{equation}
which are the solutions discovered by Meszaros \cite{Meszaros}.

The last possibility is that $\varrho=\kappa=0$ which implies $\m=1$, and we
simply obtain
\begin{equation}
\begin{aligned}
    y_1(x) &= x,\\
    y_2(x) &= x^{3/2}.
\end{aligned} \label{matter_sol}
\end{equation}

Taking into account that condition (\ref{GZ_cond}) makes it impossible for the
physical cases ($\m>0$) to have zero as a triple root, we conclude that the
only Liouvillian solutions of equation (\ref{main_eq2}) for $\L=0$, appear when
at least one of the parameters $\r$ or $\k$ is equal to zero.

\subsection{Models with the cosmological constant}

The reasoning for $\L\ne0$ is the same as in the previous section, but since
the leading coefficient now contains a polynomial of the fourth degree, the
calculations are somewhat more laborious.

In general, when $W(x)$ has only simple roots, and they are all non-zero (which
means $\r\ne 0$),
we obtain, as before, a triangular element of the monodromy group
$\mathcal M$. This leaves us with only the first case of the Kovacic's
algorithm to check, and the physical requirement of $\m>0$ causes the equation
to be non-integrable.

In fact, even if the roots become multiple, but still non-zero, the local
solutions in the vicinity of $x=0$ do not change and, again, we only need to
consider the first case of the algorithm. We introduce the roots by the
following formula
\[W(x) = \L(x-e_1)(x-e_2)(x-e_3)(x-e_4),\]
Taking $e_2=e_4$, we could rewrite the
polynomial as $W(x)=\L(x-e_1)(x+e1+2e_2)(x-e_2)^2$, because we must have
$e_1+e_2+e_3+e_4=0$. 

This time, the coefficients $b$, as defined in the
preceding section, is not numeric, but depends on the roots. We now have
\begin{equation}
    \alpha_{e_2}^{\pm} = \frac12 \pm \frac12 \sqrt{1+4\left(\frac34
    +\frac{3e_1}{e_2-e_1}-\frac{e_1}{e_1+3e_2}\right)} =: \frac{1\pm n}{2}.
\end{equation}
We see that $n$ must be at least a positive integer, and that initially there are
countably many possibilities to check. Since the relation between the roots and
the densities $\Omega$ is known, we might obtain an important restriction of
the form
\begin{equation}
    \frac{\r}{\L} = -\frac{3(n^2-4)}{n^2+12}e_2^4,
\end{equation}
which means that $\L$ must be negative if $n>2$. One can readily check that
$n=1,2$ give no solution. Further calculations with these restrictions on $\L$
and $n$, determining the coefficients of
the appropriate polynomial $P(x)$ require us to solve a system of $n$
homogeneous, linear equations. Unfortunately we have been unable to do that
for general $n$, but it is easy to check for the first 50 values that the
determinant of the system is not zero, with its modulus monotonically
increasing with $n$. Thus we conjecture that there are no values of $n$ for
which a solution exists.

When the double root becomes triple, it also becomes a pole of the third order
of the function $r(x)$. This means that only the second case of Kovacic's
algorithm needs to be considered, but it yields no Liouvillian solutions.

It is impossible for $W(x)$ to have two double roots or a quadruple root,
because as we noted the sum of all roots must be zero, and that would lead to
$\m=0$.

Letting $\r=0$ changes the multiplicity of $x=0$ as a singular point, and the
monodromy argument no longer holds. We note also that no further increase in
the multiplicity of that point is possible as $\m\ne 0$.

Assuming first that there are no multiple roots of $W(x)$, the algorithm
immediately yields the following fundamental solutions
\begin{equation}
\begin{aligned}
    y_1(x) &= x^{-3/2}\sqrt{\L x^3 +\k x +\m},\\
    y_2(x) &= y_1(x)\int\left(\frac{x}{\L x^3 +\k x +\m}\right)^{3/2}\ud x,
\end{aligned} \label{Heath_sol}
\end{equation}
which holds even if one of the roots becomes double. This solution is also
known, and was found by Heath \cite{Heath}.

As above, a triple root would mean $\m=0$ and is thus not physical.

\section{On More General Solutions}

Although the concept of Liouvillian solutions gives us a simple and applicable
formulation of solvability by quadratures, it is easily seen to be insufficient
on itself as a mean to discard an equation as insolvable in general. The fact
that the Bessel or hypergeometric functions (except for some special cases) are
not Liouvillian is the best example of that.

Of course, there are no algorithms for finding more complex solutions, but it
is possible to extend the considered class somewhat. Following the paper of
Bronstein \cite{Manuel}, we try to find solutions in the form
\begin{equation}
\begin{aligned}
    y_1(x)=m(x)F_1[\xi(x)],\\
    y_2(x)=m(x)F_2[\xi(x)], \label{spec_sol}
\end{aligned}
\end{equation}
where $F_1(\xi)$ and $F_2(\xi)$ are fundamental solutions of a given target
equation
\begin{equation}
    y''(\xi)=u(\xi)y(\xi), \label{target}
\end{equation}
and $m(x)$ and $\xi(x)$ are Liouvillian over $\mathbb C(x)$. The quoted paper
presents an algorithmic approach for target equations which have an irregular
singularity at infinity, and $\xi(x)\in\mathbb C(x)$. It also offers a check
for a certain class of algebraic $\xi(x)$.

Upon substituting the form of solutions (\ref{spec_sol}) into equation
(\ref{reduced}), and using the target equation (\ref{target}), we are left with
an expression containing only $F(\xi)$ and $F'(\xi)$. Making the coefficients
equal to zero gives the fundamental equations to be solved
\begin{gather}
    3\xi''(x)^2 - 2\xi'''(x)\xi'(x) + 4u[\xi(x)]\xi'(x)^4 - 4r(x)\xi'(x)^2
     = 0, \label{xi_eq}\\
    m(x) = \frac{1}{\sqrt{\xi'(x)}}.
\end{gather}
The main theorem of \cite{Manuel} makes it possible to find all rational
solutions of equation (\ref{xi_eq}), by bounding the degrees of the polynomials
involved.
\begin{theorem} \label{Man_th}
Let $\prod_i Q_i^i$ be the square-free decomposition of the denominator of
$r\in\mathbb C(x)$. If the order of $u(x)$ at infinity $\nu_{\infty}(u)<2$,
then any solution $\xi\in\mathbb C(x)$ of (\ref{xi_eq}) can be written as
$\xi=P/Q$ where
\begin{equation}
    Q=\prod_i Q_{(2-\nu_{\infty}(u))i+2}^i \in\mathbb C[x],
\end{equation}
and $P\in\mathbb C[x]$ is such that either
$\mathrm{deg}(P)\leqslant\mathrm{deg}(Q)+1$ or
\begin{equation}
    \mathrm{deg}(P)=\mathrm{deg}(Q)+\frac{2-\nu_{\infty}(r)}{2-\nu_{\infty}(u)}.
\end{equation}
\end{theorem}
For the algebraic case, they also provide a possible (but not exhaustive)
ansatz of the form
\begin{equation}
    \xi(x) =
    P\left(x^{1/(2-\nu_{\infty}(u))}\right)
    \prod_{i>2}Q_{i}^{(i-2)(2-\nu_{\infty}(u))}, \label{xi_alg}
\end{equation}
and a bound for $\mathrm{deg}(P)$ of either
\begin{equation}
    \mathrm{deg}(P) < (2-\nu_{\infty}(u))(\mathrm{deg}(Q)+2) \notag
\end{equation}
or
\begin{equation}
    \mathrm{deg}(P) = (2-\nu_{\infty}(u))\mathrm{deg}(Q)+2-\nu_{\infty}(r).
    \notag
\end{equation}

Here, we choose to investigate the Bessel and Kummer functions, as the
solutions of the target equation. The other classical classes of
${}_0F_1$ and ${}_1F_1$ functions, to which theorem~\ref{Man_th} is applicable,
are equivalent to these two classes, rationally or algebraically. 

In the general case of $\L\ne 0$, and no multiple roots of $W(x)$, the
denominator of $r(x)$ is $16x^2W(x)^2$, and hence its square-free decomposition
has only one term $Q_2=xW(x)$. We take first the Bessel equation,
\begin{equation}
    F''(\xi) = \left(\frac{4n^2-1}{4\xi^2}-\epsilon\right)F(\xi),
\end{equation}
where $\epsilon$ is 1 for the Bessel and -1 for modified Bessel functions. We
have $\nu_{\infty}(u)=0$, and $\nu_{\infty}(r)=2$, so
\begin{equation}
    Q = \prod_i Q_{2i+2}^i = 1,
\end{equation}
and $\mathrm{deg}(P) \leqslant 1$. Finally, substituting $\xi(x)=c_1 x+c_0$ into
equation (\ref{xi_eq}) yields no non-constant solutions.

Proceeding in the same way, for Kummer functions, algebraic forms of $\xi(x)$,
and possible cases of the roots of $W(x)$, we found no new solutions of this
more general class.

Thus, we can finally formulate the main results as the following theorem:
\begin{theorem}
If $\m>0$, $\r\geqslant 0$, $\k\in\mathbb R$, and $\L\in\mathbb R$ (unless
$W(x)$ admits a double root, which requires the assumption of $\L\geqslant 0$),
then the matter density perturbation equation (\ref{main_eq2}) does not posses
any solutions Liouvillian over $\mathbb C(x)$, or, in other words, is not
solvable by quadratures, except for the following three cases:
\begin{enumerate}
\item Heath's solution ($\r=0$). It is given by (\ref{Heath_sol}), for
$\L\ne 0$, and by (\ref{Heath_0}), when $\L=0$.
\item Meszaros's solution ($\k=0$, $\L=0$). Given by (\ref{flat_sol}).
\item Matter only ($\L=\r=\k=0,\m=1$). Given by (\ref{matter_sol}).
\end{enumerate}
Moreover, there exist no solutions of the class $m(x)F[\xi(x)]$,
where $F(\xi)$ is a classical special function of the type ${}_0F_1$ or
${}_1F_1$, $m(x)$ is Liouvillian over $\mathbb C(x)$,
and $\xi(x)\in\mathbb C(x)$ or is algebraic of the form (\ref{xi_alg}).

However, when $\L=0$ and $W(x)$ admits a double root, the non-Liouvillian
solution of Chernin's, given by (\ref{Chernin_sol}), applies.
\end{theorem}

\section{Conclusions}

The problem studied in this paper is a good example of a possible application
of the differential Galois theory in practise. The known solution of the
perturbation equation were originally discovered in different contexts and over
the span of a few years, whereas here, one theory allows for full analysis.

Furthermore, the existence of other closed-form solutions has
been ruled out. The one exception is the case of a double root of $W(x)$, when
the cosmological constant is negative. Although $\Lambda$ is usually assumed to
be positive now, it still remains a viable mathematical possibility. However,
we expect that that case will not yield any new solutions, as the ``manual''
check for the first, simplest candidates failed.

Despite the fact that the analysis of perturbations can be performed
numerically, without the need of explicit solutions, we feel that a more exact
approach is always valuable -- providing a better understanding of the
complexity of the problem, revealing hidden, special solutions, or simply
providing a more exact formula for numerical work. It also reflects, in
some sense, the fact that general relativity, as the next step in the theory of
gravitation, introduces non-integrability into the basic equations. This can be
seen clearly in a more advanced employment of the Galois theory applied to
cosmological models themselves, for example in \cite{Morales} and
\cite{Maciejewski}.

From the physical point of view, this model is a relatively simple one, and it
is already known that the solutions found are not on themselves strong enough
to explain the density fluctuations levels today. The main problem is the fact
that they would need to be exponential, whereas they grow at most like a fixed
power of the scale factor,
and there is not enough time for the initial perturbations to increase
sufficiently. One has also to be aware that linear instability might be dumped
when considered in the quadratic regime, so the least to be required of the
linear equation's solutions is appropriate amplification. Nevertheless, these
basic effects, being the first approximation, have physical meaning and should
be taken into account when constructing more complicated models or explaining
observational data.

We hope to investigate this field further in the future, as it does not limit
the results to negative statements on integrability only, as we tried to
demonstrate in this paper.

\section*{Acknowledgements}
T.S. has been supported by the KBN grant number 2P03D 00 326.

\end{document}